\documentclass[a4paper,12pt]{amsart}
\usepackage[utf8]{inputenc}
\usepackage{geometry}
\usepackage{setspace}
\newtheorem{thm}{Theorem}[section]

\newtheorem{lem}[thm]{Lemma}

\usepackage{graphicx}
\newcommand{\VV}{V_{\rm vis}}
\newcommand{\II}{\mathbb I}
\newcommand{\CC}{{\rm Cov}}
\newcommand{\bN}{\mathbb{N}}
\newcommand{\NN}{\widetilde{N}}

\usepackage{color}

\newcommand{\blue}{\textcolor{black}}
\begin{document}

\title{Normalising phylogenetic networks}
\author{Andrew Francis, Daniel H. Huson and Mike Steel}
\date{\today}

\doublespace
\begin{abstract}
Rooted phylogenetic networks provide a way to describe species' relationships when evolution departs from the simple model of a tree. However, networks inferred from genomic data can be highly tangled, making it difficult to discern the main reticulation signals present. In this paper, we describe a natural way to transform any rooted phylogenetic network into a  simpler canonical network, which has desirable mathematical and computational properties, and is based only on the `visible' vertices in the original network. The method has been implemented and we demonstrate its application to some examples. 
\end{abstract}

\maketitle
\singlespace

\noindent {\em Address:}
\\
             Andrew Francis\\
Centre for Research in Mathematics and Data Science \\
Western Sydney University, Australia
\\
\email{A.Francis@westernsydney.edu.au}  \\
\\
              Daniel H. Huson\\
              Algorithmen der Bioinformatik \\
              Universit{\"a}t T{\"u}bingen, Germany
              \\
 \email{daniel.huson@uni-tuebingen.de}  
              \\
              \\
           M. Steel (corresponding author):\\
              Biomathematics Research Centre, \\University of Canterbury, Christchurch, New Zealand \\
              Tel.: +64-3-364-2987 ext 7688\\
              Fax: +64-3-364-2587\\
              \email{mike.steel@canterbury.ac.nz}  
 \\
 \\

\doublespace
\newpage

\section{Introduction}

Rooted phylogenetic trees are the most widely used means to represent  evolutionary relationships among species, under simple processes of speciation and extinction.  However, phylogenetic trees have their limitations: when reticulate processes such as hybridisation or lateral gene transfer have also been involved, rooted phylogenetic networks provide a more accurate and complete representation of evolutionary history.  Various methods exist for building phylogenetic networks, and these networks can sometimes be very complex and tangled \cite{tal09}. Unlike phylogenetic trees, where the number of interior vertices and edges  is bounded above (linearly) by the number of leaves, phylogenetic networks can have an unbounded number of interior vertices and edges.  Thus, it is useful to try to summarise a phylogenetic network with a simpler graphical structure, that still captures the most important features. 

One option is to summarise a network by a tree that represents some central tendency or underlying tree pattern  \cite{dre10, koo}. However, such an approach discards all the reticulation signals; moreover, the resulting tree is often poorly resolved. 

In this paper, we describe the construction of a canonical `normal' network $\NN$ associated with any rooted phylogenetic network $N$ (the word canonical here refers to the absence of any arbitrary choices in the construction of $\NN$ from $N$).  We  call $\NN$ the `normalisation' of $N$, and it is structurally and computationally tractable.  \blue{The construction of $\NN$ focuses on a subset of the vertices of $N$ --- the `visible' vertices ---  that are unavoidable in any path from the root to at least one extant species, with any resulting `short-cut' arcs subsequently removed. } 

Our approach is consistent with the philosophy  that it is futile to try to distinguish among phylogenetic networks that are essentially non-distinguishable from the data available at the present (as articulated in \cite{par15}, in a  setting where the data consisted of \blue{trees with edge lengths}). \blue{ Some aspects of our approach relate to earlier work: In \cite{mor04}, a procedure for simplifying certain types of networks (called `reconstructible networks') was described (Section 4.4 of that paper), and in \cite{wil11} an approach for reconstructing normal networks directly from rooted trees was presented and discussed.}

The structure of the paper is as follows. 
We begin by recalling some key definitions concerning phylogenetic networks, and describe classes of networks and their relationships to each other. In Section~\ref{sec:nor}, we formally define the normalisation of a network and illustrate it with a simple example.  In Section~\ref{sec:pro}, we establish the main properties of this construction (Theorem~\ref{thm1}), which also provides a characterisation for the types of networks that have a tree as their normalisation. We then discuss the implications of this construction, provide a short application to a previous data set, and end with some brief concluding comments. 

\subsection{Definitions}
A (rooted) {\em phylogenetic network} for a set $X$ of species, is a finite acylic directed graph $N=(V,A)$ with a single vertex of in-degree 0 (called the {\em root} and denoted $\rho$) and for which the remaining vertices in $V$ fall into three disjoint classes: 
\begin{itemize}
\item[(1)] vertices of in-degree 1 and out-degree 0 (the {\em leaf set} of $N$), which comprise the set $X$;
\item[(2)] vertices of in-degree 1 and out-degree at least 2;
\item[(3)] vertices of out-degree~1 and  in-degree at least 2.
\end{itemize}
\blue{In biology, the leaf set $X$ generally corresponds to the set of extant species (or taxa) under study, and vertices of type (2) describe speciation events, while vertices of type (3) describe events involving reticulate evolution (e.g. hybridization, lateral gene transfer, endosymbiosis). This latter class of vertices are referred to as {\em reticulate vertices}.}

We let $RPN(X)$ denote the set of rooted phylogenetic networks on leaf set $X$ up to equivalence (two networks are regarded as equivalent if there is a digraph isomorphism between them that maps leaf $x$ in the first network to leaf $x$ in the second, for all $x \in X$). Further details on phylogenetic networks can be found in  \cite{hus10} and in Chapter 10 of \cite{ste}.

For $N\in RPN(X)$ and two vertices $u,v\in V(N)$, write $u\xrightarrow[N]{} v$
if there is a path from $u$ to $v$ in $N$.  The {\em interior} of a path is the set of vertices in the path without the endpoints included, \blue{and a vertex is {\em interior} if it is not a leaf vertex. For an arc $(u,v)$ in a network, $u$ is said to be the {\em parent} of $v$ and $v$  the {\em child} of $u$.}
We will say that a vertex in a directed graph is {\em subdividing} if its in-degree and out-degree both equal 1.
\blue{Note that networks in $RPN(X)$ have, by definition, no subdividing vertices.} 

A vertex $v$ in $N$ is said to be {\em visible} (in $N$) if there is a leaf $x \in X$  so that every path from the root vertex $\rho$ of $N$ to leaf $x$ includes $v$. 
The biological relevance of visibility arises from the desire to
reconstruct evolutionary history from \blue{genomic data observed at the present} (i.e. from the genomes of the species in the set $X$); \blue{ if a vertex $v'$ in a phylogenetic network is not visible, then the evolutionary pathway carrying genomic information\footnote{\blue{Here, `genomic information'  refers to any changes in ancestral genomes  that can be passed from an ancestral species to a descendent species (for example, genomic insertions, deletions, rearrangements, mutations).}} from the root down to the extant species may have simply bypassed vertex $v'$.  The visibility condition is nevertheless quite strong, as it requires that there is a leaf for which {\em every} path from the root to that leaf goes through $v’$. }

We will let $\VV(N)$ denote the set of visible vertices of $N$. 
Note that $\VV(N)$ always includes the set $X$ of leaves of $N$, as well as the root of $N$.
For \blue{an interior} vertex $v$, let $\II(v)$ be the set of leaves $x$ with the property that $\rho$ and $x$ are disconnected in the digraph obtained from $N$ by deleting $v$ and its incident arcs.  Thus \blue{an interior} vertex $v$ is visible if and only if $\II(v) \neq \emptyset$.
We say that any leaf in $\II(v)$ {\em identifies} $v$ in $N$.

 For $v\in V(N)$,  let $C(v)$ denote the set of leaves descended from $v$ \blue{(i.e. the set of leaves $x \in X$ with $v\xrightarrow[N]{} x$)} referred to as the {\em cluster} corresponding to $v$.  Clearly,  $\II(v)\subseteq C(v)$ for each \blue{interior vertex} $v$, and $\II(v)=C(v)$ for \blue{every interior vertex} $v$ if $N$ is a tree.

\subsection{Subclasses of networks}

Apart from trees, there are three other subclasses of $RPN(X)$ that are relevant to this paper. Each of these classes can be defined in various (equivalent) ways; here, we have chosen the simplest or most relevant definition.  

\begin{figure}[h]
   \includegraphics[scale=0.4]{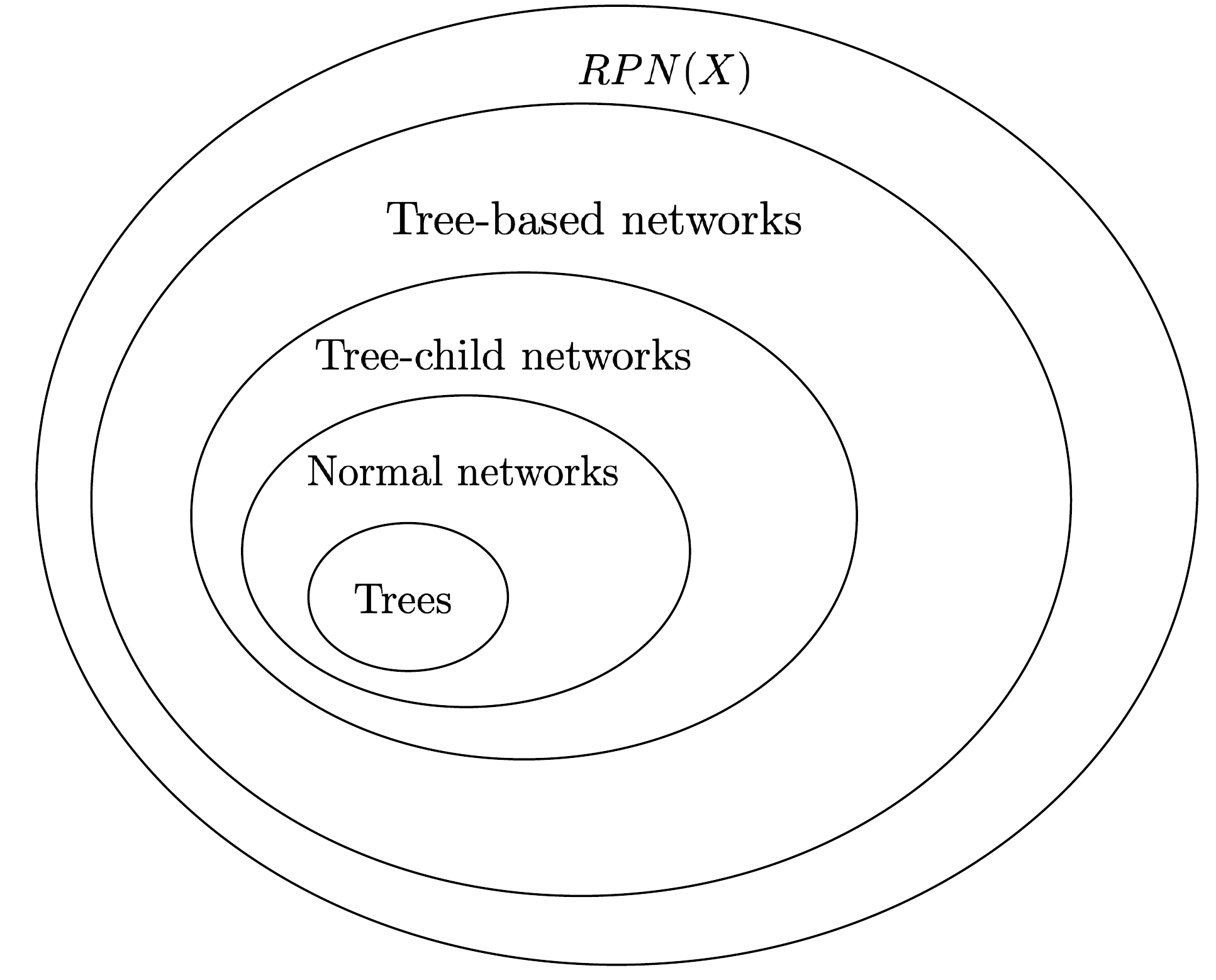}
   \caption{ The nested relationships of the sub-classes of networks described.}
      \label{fig1}
    \end{figure}

We begin with the class of `tree-child' phylogenetic networks, introduced by Cardona et al. \cite{car09}. A network is said to be {\em tree-child} precisely if every vertex is visible (i.e. $\II(v) \neq \emptyset$ for \blue{each interior vertex} $v \in V$). A network is {\em normal} if it is a tree-child network that has the additional property of containing no `shortcuts' (i.e. arcs $(u,v)$ for which there is already a path from $u$ to $v$)~\cite{willson2010properties}. 
A network is {\em tree-based} if it has a rooted spanning tree with leaf set $X$~\cite{francis2015phylogenetic}.
The nesting of these various classes of networks is indicated in Fig.~\ref{fig1} \blue{(for further details, see \cite{ste}, p.261).}

\section{The normalisation of a network}
\label{sec:nor}

Given $N$, consider the partial order $\leq_N$ on the set $\VV(N)$ of visible vertices of $N$ defined by $v\leq_N v'$
if $v\xrightarrow[N]{} v'$.  \blue{In this way, we can associate to $N$ another directed graph  $\CC(N)$ with vertex set $\VV(N)$ as follows:
For every pair of vertices $u, v \in \VV(N)$ for which there is a directed path in $N$ from $u$ to $v$ place an arc from $u$ to $v$, and once all such arcs have been inserted then remove any shortcut arcs  (the graph $\CC(N)$ thus corresponds to the ``Hasse diagram" of $\VV(N)$ under the partial order $\leq_N$). 
Note that $\CC(N)$ may have subdividing vertices so we will let $\NN$ denote the network obtained from $\CC(N)$ by suppressing each  subdividing vertex.} We call $\tilde N$ the \textit{normalisation} of $N$, and write  $\varphi: RPN(X) \rightarrow RPN(X)$ for the normalisation function $N \mapsto \NN.$

\bigskip

\begin{figure}[h]
   \includegraphics[scale=0.8]{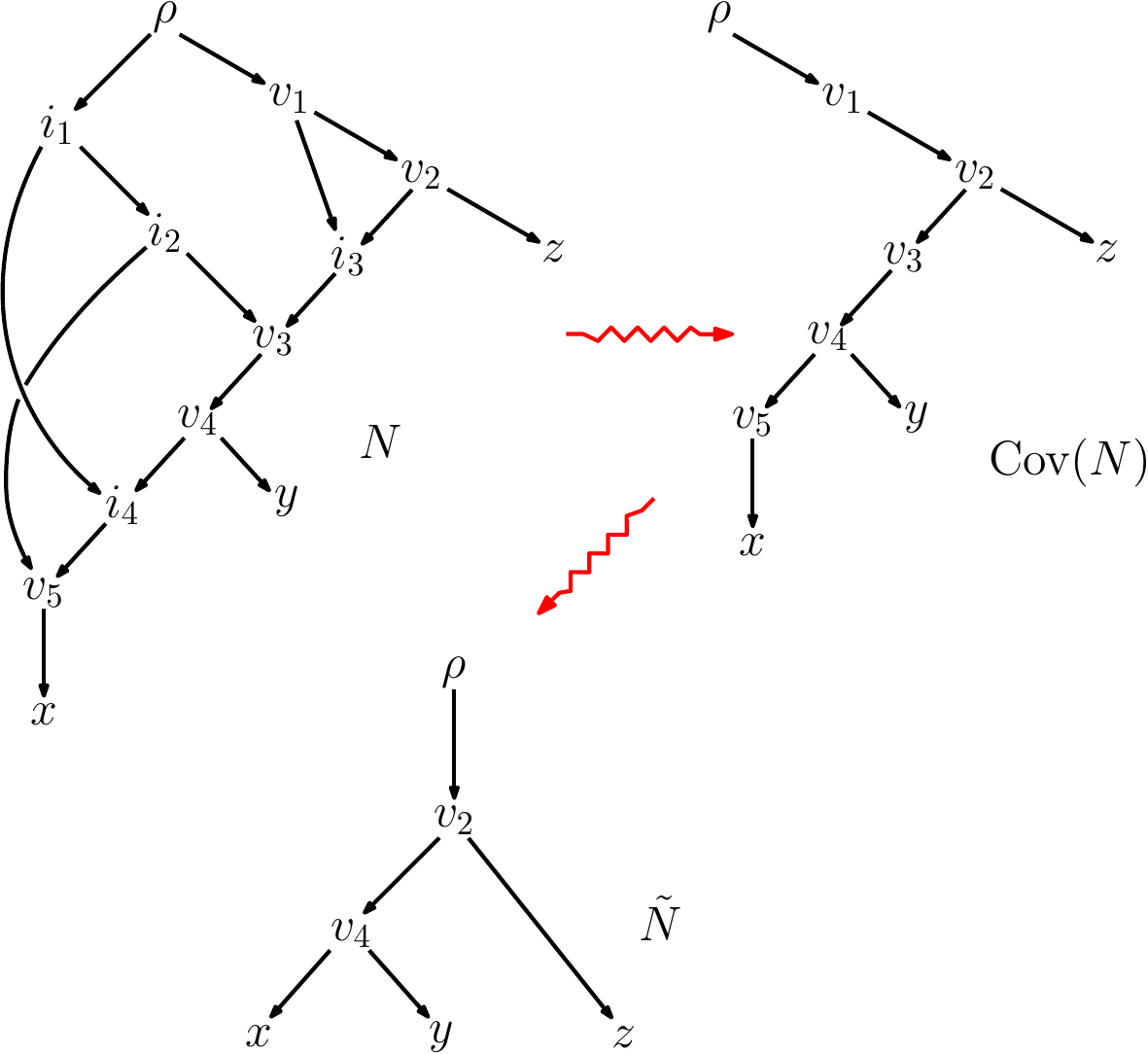}
   \caption{The normalisation of the non-tree-based network $N$, produces a tree ($\NN$). Vertices labelled $i_*$  in $N$ are not visible in $N$; the other vertices in $N$ (labelled $v_*$, together with the root $\rho$ and the leaves $x,y,z$) are visible. }
      \label{fig2}
    \end{figure}

Note that the root of $\NN$ may have out-degree 1, even if the root of $N$ has a higher out-degree (as in  Fig.~\ref{fig2}). Notice also that the vertex set $V(\NN)$ of $\NN$ is a subset of $\VV(N)$, and the inclusion can be strict because in moving from $\CC(N)$ to $\NN$,  subdividing vertices that are visible are suppressed.

An example is provided in Fig.~\ref{fig2}, which involves a network from Fig 3 of ~\cite{francis2015phylogenetic} that is not tree-based (but satisfies a certain `antichain-to-leaf' property). 

\section{Properties of the normalisation construction}

\label{sec:pro}
In order to state our main result, we first need to establish some basic properties, summarised in two lemmas.  

\begin{lem}
\label{lem2}
Suppose that there is a path $p$ from $u$ to $v$ in $\NN$. In that case, there is a path in $N$ from $u$ to $v$ that includes all the vertices in $p$ (and possibly additional vertices).
\end{lem}
\begin{proof}
Suppose that $(u,v)$ is an arc in $\NN$. There is then a path in $N$ from $u$ to $v$, by the definition of  $\CC(N)$. Repeated application of this argument establishes the lemma.
\end{proof}

The proof of the second lemma is provided in the Appendix. 

\begin{lem}
\label{lem1}

\mbox{}
\begin{itemize}
    \item[(a)] \blue{The normalisation network $\NN$} has no `shortcuts' (arcs $(u,v)$ for which there is already a path from $u$ to $v$). 
    \item[(b)] For any network $N \in RPN(X)$, the normalisation network $\NN$ lies in $RPN(X)$.
\end{itemize}
\end{lem}

We can now state our main result, the proof of which is provided in the Appendix. For Part (iv), recall that a {\em hierarchy} on $X$ is a collection of subsets of $X$ that satisfies the property that any two sets are either disjoint, or one is a subset of the other. 

\begin{thm}
\label{thm1}
\mbox{}
\begin{itemize}
    \item[(i)]
For any $N\in RPN(X)$, $\tilde N$ is a normal network in $RPN(X)$. Moreover, 
    $\NN = N$ if and only if $N$ is a normal network. \blue{Moreover,} the normalisation function $\varphi$ is idempotent (i.e. $\widetilde{\NN}=\NN$).

\item[(ii)]
For any two vertices $u, v$ of $\NN$, $u \leq_{\NN}v \mbox{ if and only if }u \leq_N v.$
\item[(iii)] 
Let $v$ be a vertex of $\NN$. The cluster $C_{\NN}(v)$ is then identical to the cluster $C_{N}(v)$.

\item[(iv)] 
$\NN$ is a tree if and only if $\{C_N(v): v \in \VV(N)\}$ forms a hierarchy on $X$.

\end{itemize}
\end{thm}

\subsection{Application}
\begin{figure}[htbp]
   \includegraphics[scale=1.0]{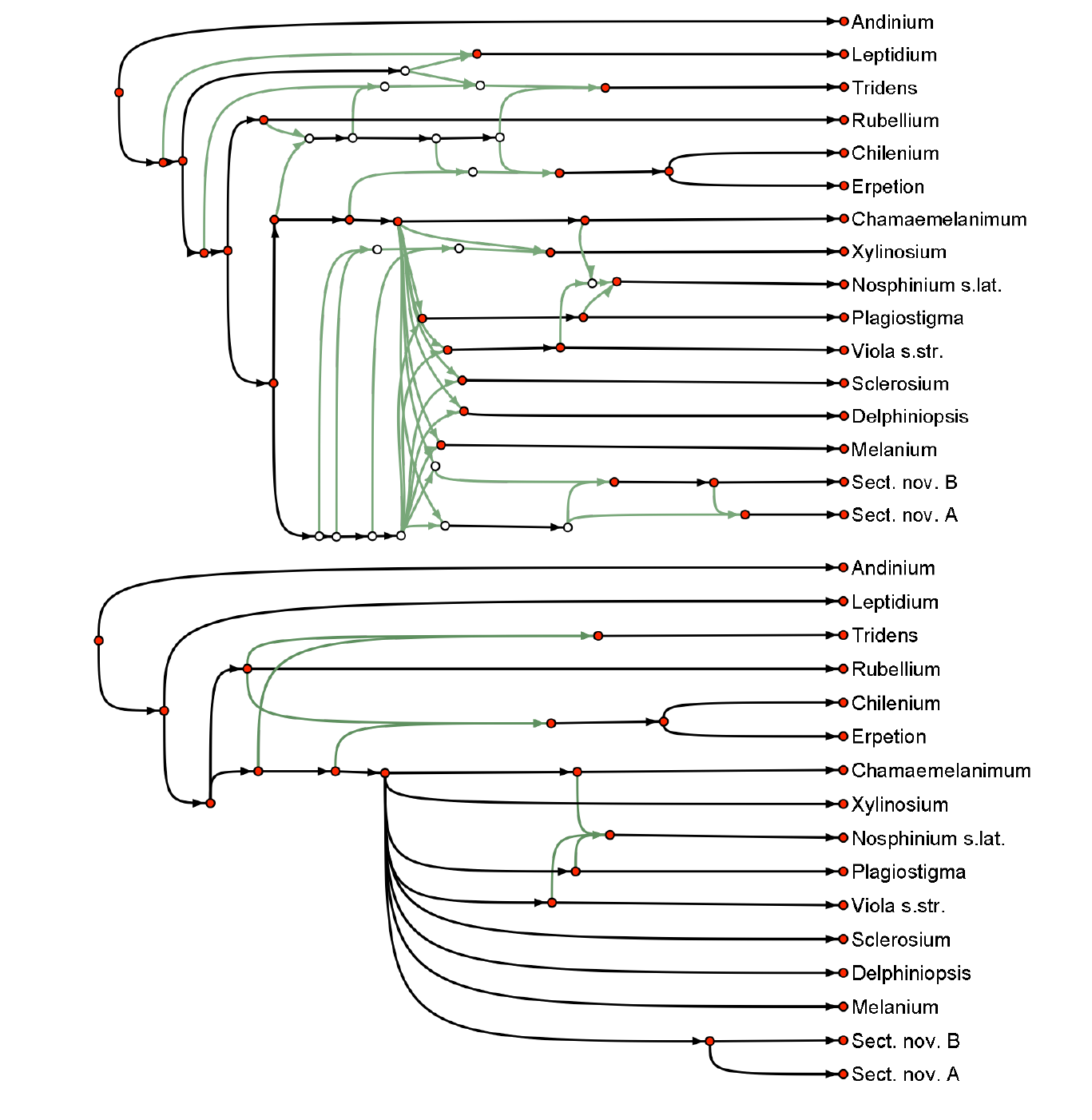}
   \caption{Top: A rooted phylogenetic network $N$ from \cite{jet18} based on a study from \cite{mar15} \blue{with the visible vertices colored red.}  Bottom: The normalisation $\NN$ of $N$. The labelling of the leaves in $\NN$ (from top to bottom) matches that in $N$.}
      \label{fig3}
    \end{figure}

 In Fig.~\ref{fig3}, we illustrate the computation of the normalisation of a biological network involving reticulate evolution in a study of the {\em Viola} genus from \cite{mar15}. This network $N$ was investigated in \cite{jet18} as an example of a network for which the pattern of reticulation is such that the original network is not even tree-based. The network $\NN$ is produced using {\em PhyloSketch} \cite{phylo}. 

\newpage

{\bf Remarks}
\begin{enumerate}
 
\item 
\blue{ The network $\NN$ can be constructed quickly (i.e. in polynomial time) from $N$, making it applicable for large rooted networks (binary or non-binary),  even when these contain thousands of vertices and edges. An  algorithm for constructing the normalisation of a network has been implemented in an open-source program called {\em PhyloSketch} \cite{phylo}. In the Appendix we provide some further information on this algorithm and its running time.}

    \item
     If $\bN(X) \subseteq RPN(X)$ denotes the set of normal networks on $X$, then Theorem~\ref{thm1}(i) states that the normalisation function  $\varphi$ can be viewed as a type of retraction from $RPN(X)$ onto $\bN(X)$.
However, the map $N\mapsto \NN$ is not a connected surjective digraph map (CSD) in the sense of 
Willson~\cite{Willson2012csd} because the edges in $\NN$ are not a subset of the edges in $N$.

\item  
    
Since every normal network is tree-based, the map $N \mapsto \NN$  associates a canonical tree-based network to any network. Moreover, since normal networks are of the tree-child type and are thus stack-free (i.e. no reticulate vertex has a reticulate parent) it follows that if all reticulations in $\NN$ have in-degree 2, there always exists at most two trees that cover every edge of $\NN$ (due to a result in \cite{sem18}).

\item 
The condition that the set
 $\{C_N(v): \blue{v \in \VV(N)}\}$ forms a hierarchy on $X$ does not imply that $N$ is a tree (Fig.~\ref{fig2}  provides a counterexample); however, this condition suffices for $\NN$ to be a tree (by Part (iii) of the previous theorem, since any subset of a hierarchy is a hierarchy).  On the other hand, the condition that $\{\II(v): \blue{v \mbox{ is an interior vertex of } N } \}$ forms a hierarchy does not suffice for $\NN$ to be a tree; an example illustrating this is provided in Fig.~\ref{fig4}.
 Moreover, in this example, the network $\NN$ is not  displayed\footnote{Given two networks $N, N' \in RPN(X)$
with $V(N') \subseteq V(N)$,
we say that $N$ {\em displays} $N'$ if (i) for each arc $(u,v)$ of $N'$, there is an associated path $p(u,v)$ in $N$ (consisting of at least one arc) from $u$ to $v$; and (ii) for any two arcs $(u,v)$ and $(u', v')$, there is no vertex of $N$ common to to the interior of both $p(u,v)$ and $p(u',v')$.}  by $N$. \blue{Remarkably, the set of trees displayed by $N$  in this example and the set of trees displayed by $\NN$ share no tree in common.}

  \begin{figure}[htbp]
   \includegraphics[scale=1.0]{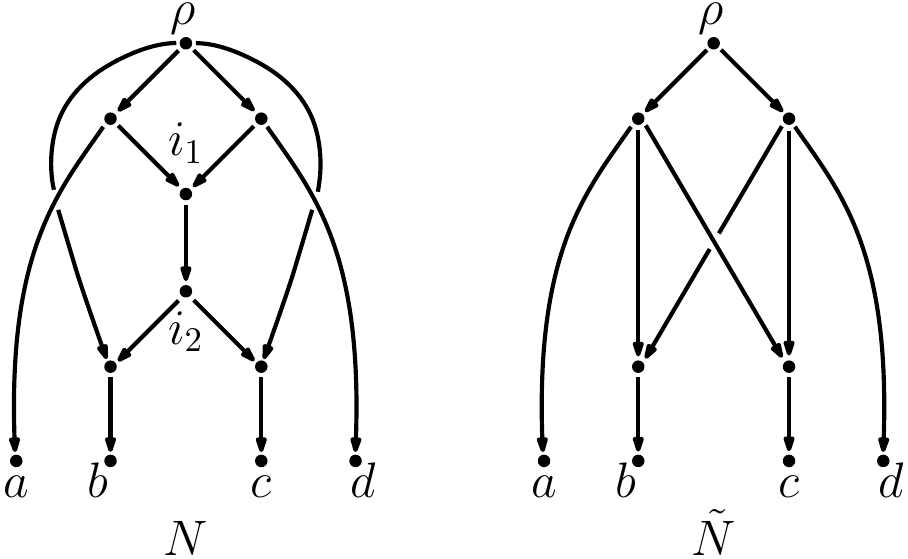}
   \caption{The network $N$ (left) has all its vertices visible except for vertices $i_1$ and $i_2$. This network has the property that $\{\II(v): \blue{v \mbox{ is an interior vertex of } N } \}$ is a hierarchy on $X$, yet $\NN$ is not a tree.}
      \label{fig4}
    \end{figure}
    
 \item
 Since $\NN$ is a tree-child network, $\NN$ has at most $n-1$ reticulations where $n=|X|$ by a result in \cite{car09}; in fact, since $\NN$ is normal, this bound improves to $n-2$ (\cite{ste}, p. 252).  Moreover, if  $\NN$ is also binary then it displays exactly $2^r$ trees (by  Corollary 3.4 of \cite{wil12}, which provides a more general result in the case where $\NN$ is non-binary).
  \item 
    One can define two binary operations $\oplus$ and $\otimes$ on rooted phylogenetic networks as follows: $N_1 \oplus N_2$ is the phylogenetic network consisting of disjoint copies of $N_1$ and $N_2$ (which have disjoint leaf sets) whose roots are incident with a new root vertex, and $N_1 \otimes N_2$ is the phylogenetic network obtained by identifying each leaf of $N_1$ with the root of a copy of $N_2$ (again with disjoint leaf sets).  The operation $\oplus$ is commutative but not associative, whereas $\otimes$ is associative but not commutative.
 Moreover, the distributive law also applies:  $$(M \oplus M') \otimes N = (M \otimes N) \oplus (M' \otimes N).$$ 
  It can be checked that $N \mapsto \NN$ respects these operations. In other words:  
   $\varphi(N\oplus N')= \varphi(N) \oplus \varphi (N') \mbox{ and } 
   \varphi(N\otimes N')= \varphi(N) \otimes \varphi (N').$
   
\end{enumerate}

\bigskip

%This gives a lower bound on the number of trees displayed by $N$, using the following lemma.

\section{\blue{Discussion and } concluding comments}

\begin{figure}[htbp]
   \includegraphics[scale=1.0]{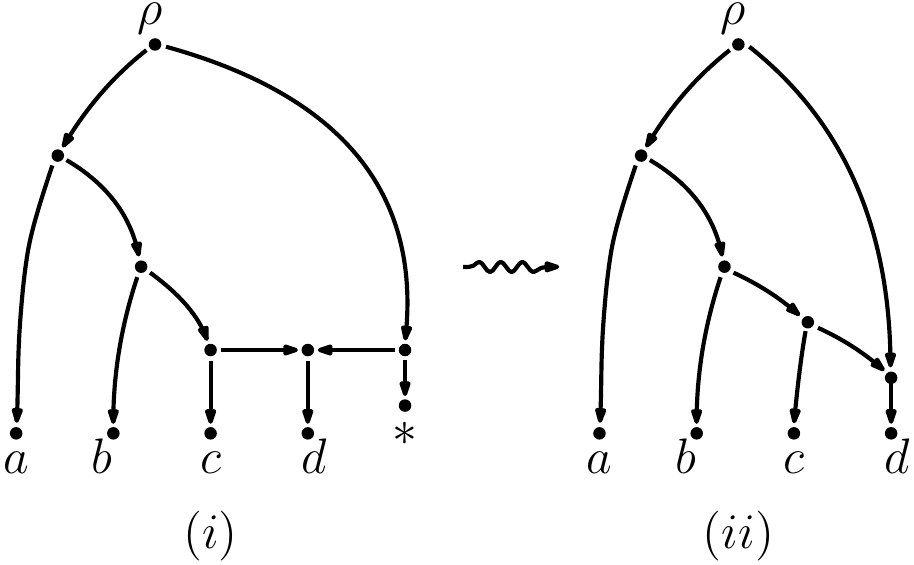}
   \caption{
   \blue{(i) An evolutionary  scenario involving extant species $a, b, c, d$, and a species ($*$) that is either extinct or unsampled.
Species $d$ is the child of a reticulate vertex that corresponds to a hybridization between a parent of $*$ and a parent of $c$.  (ii) The  induced network on the species $a,b,c,d$ contains a `shortcut' arc. } 
  }
           \label{fig5}
    \end{figure}

\blue{Since the normalisation function can transform an arbitrarily complex phylogenetic network into a relatively simple network, it is important to note that underlying evolutionary pathways may sometimes be lost in making this transition. Fig.~\ref{fig5} illustrates an example of how this can occur when an ancestral hybridization event involves a lineage that leads to an extinct or un-sampled species (denoted by $*$ in the figure). This  results in a phylogenetic network on the remaining extant species ($a, b, c, d$) for which a full representation of all evolutionary pathways  (shown in Part (ii) of the figure)  requires a `shortcut' arc (from the root vertex $\rho$ to the parent of $d$) and this arc would be deleted in the normalisation step.  However, some cases where there is a loss of information regarding the evolutionary history of extant taxa is to be expected in any procedure that simplifies an arbitrary phylogenetic network. }

\blue{Turning next to future directions,} Theorem~\ref{thm1} suggests a natural question: For a given normal network $N'$, what  properties do networks that normalise to $N'$ have?  When $N'$ is a tree, Theorem~\ref{thm1}(iii) provides a precise answer. However, more generally, for a general $N'$ what can we say about the preimage $\varphi^{-1}(N')$?  That is, can one characterise the set
$\{N\in RPN(X): \varphi(N)= N'\}$?
This question suggests a natural equivalence relation on networks: $N_1\cong N_2$ if and only if $\NN_1\equiv\NN_2$ (phylogenetic network isomorphism).  

The  question above is basically asking how we can describe these equivalence classes of network. One feature that is easy to see is that any such equivalence class of a non-trivial network will have infinitely many networks in it, i.e., $|\{N\in RPN(X): \varphi(N)= N'\}|=\infty$.  This can be seen as follows.  Take any network in the fiber (the equivalence class), and select any edge in the network.  Subdivide it into 5 parts with new vertices $a,b,c,d$; add edges  $(a,c)$ and $(b,d)$; now add edges `parallel' to $(b,d)$ from the edge above $b$ to the edge below $d$, as many as you like.  No visibility of vertices from the original network is changed, and the normalization clearly gives the same network.  

A further question is how the approach described in this paper might be extended to deal with {\em unrooted} phylogenetic networks.

 \section{Funding}
 We thank the Royal Society Te Ap\={a}rangi (New Zealand) for funding under the Catalyst Leader program (Agreement \# ILF-UOC1901). 
   
 \blue{\section{Acknowledgements} We thank the reviewers and handling editor for a number of helpful comments and suggestions on an earlier version of this manuscript.}

\bibliographystyle{plain}
\bibliography{francis_huson_steel_revision}

\newpage

\appendix

\section{Proof of Lemma~\ref{lem1} and Theorem~\ref{thm1}.}

\subsubsection*{Proof of Lemma~\ref{lem1}}\ 
% {\em Proof of Lemma~\ref{lem1}}.

{\em Part (a):} 
Let $N'$ be the network obtained by suppressing all subdividing vertices in $\CC(N)$, and let $(u,v)$ be an edge in $N'$.  Let $p$ be a path from $u$ to $v$ of length $>1$ in $N'$.  We will derive a contradiction. 

Note that $(u,v)$ cannot be an edge of $\CC(N)$ otherwise $\CC(N)$ would have a shortcut  \blue{(and this would contradict the definition of the graph $\CC(N)$).} Thus, $(u,v)$ must correspond in $\CC(N)$ to a path $p'$ from $u$ to $v$ the interior of which consists only of (one or more) subdivision vertices. 

Select the subdivision vertex on $p'$  that is closest to $v$ and call it $w$.   Since $w$ is a visible vertex in $N$, it cannot be a reticulate vertex, otherwise the path $p$ would provide a path between the root of $N$ and the leaves descended from $w$ but which avoids $w$.   Thus, there are arcs $(w, w_1), \ldots, (w, w_k)$ ($k\geq 1$) in $N$, where $w_i$ are visible vertices of $N$. Moreover, since $w$ is a subdivision vertex of $\CC(N)$, for each such arc $(w,w_i)$ there is (at least) one path $p_i$ in $N$ from $v$  to $w_i$ (else the arc $(w, w_i)$ would appear in $\CC(N)$ and so $w$ would not be a subdivision vertex).  It follows that $w$ is not a visible vertex of $N$, since any path from a leaf to the root that goes via $w$ can be avoided by using the paths $p_i$ and $p$. This establishes the required contradiction.

\bigskip 

{\em  Part (b):} First, observe the following:
\begin{itemize}
\item[(i)] $\NN$ has the same root vertex as $N$ (having in-degree 0 in $\NN$) and each element of $X$ is a leaf of $\NN$ (i.e. the root and leaves of $X$ are always visible vertices and not subdividing);
\item[(ii)]  $\NN$ contains no vertex of out-degree 0 that is not in $X$ and no vertex of in-degree 0 except the root;
\item[(iii)] $\NN$ has no subdividing vertices (since these have been removed in the construction). 
\end{itemize}
There is one further condition to check to ensure that $\NN \in RPN(X)$: 
\begin{itemize}
\item[(iv)]
$\NN$ has no vertex of in-degree and out-degree both at least $2$. 
\end{itemize}
To establish Part (iv), suppose that $v$ is a vertex of $\NN$ for which the  in-degree and out-degree of $v$ are both at least 2 (we will derive a contradiction). Let $x_v$ be a leaf in $\II(v)$.  In $N$, the vertex $v$ either has out-degree 1 (Case (a)) or in-degree 1 (Case (b)). In Case (a), let $w$ denote the (unique) child of $v$; in Case (b), let $w$ denote the (unique) parent of $v$. In either case,  $w$ is a visible vertex of $N$ (since $x_v \in \II(w)$).
Thus, in $\CC(N)$, $v$ has out-degree 1 (in Case (a)) or out-degree 2 (in Case (b)) and suppressing any subdividing vertices in converting $\CC(N)$ to $\NN$ does not alter this conclusion, thereby contradicting the assumption regarding $v$.
\hfill$\Box$

% {\em Proof of Theorem~\ref{thm1}.}

\subsubsection*{Proof of Theorem~\ref{thm1}}\ 
% \begin{itemize}
% \item[(i)] 
    % For any $N\in RPN(X)$, $\tilde N$ is a normal network in $RPN(X)$. 
    
    % Moreover, $\NN = N$ if and only if $N$ is a normal network, and the normalisation function $\varphi$ is idempotent (i.e. $\widetilde{\NN}=\NN$).

% \item[(ii)] 
    % For any two vertices $u, v$ of $\NN$, $u \leq_{\NN}v \mbox{ if and only if }u \leq_N v.$

% \item[(iii)] 
    % Let $v$ be a vertex of $\NN$. The cluster $C_{\NN}(v)$ is then identical to the cluster $C_{N}(v)$.

% \item[(iv)] 
    % $\NN$ is a tree if and only if $\{C_N(v): v \in \VV(N)\}$ forms a hierarchy on $X$.
% \end{itemize}

{\em Part (i):}  First we establish that each vertex $ v$ of $\NN$ is visible in $\NN$.  
Since $v$ is visible in $N$ (by definition of the vertex set of $\NN$), there is a leaf $x = x_v$ for which each path from $\rho$ to $x$ includes $v$. 
Let $p$ be any path in $\NN$ from $\rho$ to leaf $x$.   By Lemma~\ref{lem2}, $p$ corresponds to a path $p'$ in $N$ from $\rho$ to $x$. Since $v$ is visible, $p'$ includes the vertex $v$.  This holds for all choices of $p$,  so v is a visible vertex in $\NN$, as claimed. 

Since all vertices of  $\NN$ are visible, this network is tree-child.  Furthermore, by Lemma~\ref{lem1}, $\NN$ lies in $RPN(X)$ and has no shortcuts, so it is a normal network in $RPN(X)$.

{\em Part (ii):}  The implication $\Rightarrow$ follows from Lemma~\ref{lem2}.  For the converse, suppose that $u \leq_N v$ for $u,v$. Then there is a path from $u$ to $v$ in $N$, and so there is a path in ${\rm Cov}(N)$ from $u$ to $v$. Suppressing any subdividing vertices does not eliminate any path from $u$ to $v$.

{\em Part (iii):} 
Suppose that $x \in C_{\NN}(v)$. There is then a path in $\NN$ from $v$ to $x$ and so, by  Part (ii), there is a path from $v$ to $x$ in $N$, and therefore $x \in C_{N}(v)$.  The same argument applies if $x \in C_{N}(v)$ (by the other direction in Part (ii)). 

{\em Part (iv):}  By Part (iii), we have: 
\begin{equation}
\label{clusterseteq} 
\{C_{\NN}(v): v \in V(\NN)\} = \{C_{N}(v): v \in V(\NN)\}.
\end{equation}  Moreover,  regarding the set on the RHS of this equality we have:
\begin{equation}
\label{clusterseteq2} 
\{C_{N}(v): v\in V(\NN)\} = \{C_{N}(v): v\in V(Cov(N))\},
\end{equation}
since a vertex of in-degree and out-degree 1 has the same cluster as its child (i.e. such vertices do not introduce new clusters, nor are clusters lost when such vertices are removed in the construction of $\NN$ from $Cov(N)$). 
If we observe that $V(Cov(N))=\VV(N)$ (the set of visible vertices in $N$), then from Eqns.~(\ref{clusterseteq}) and ~(\ref{clusterseteq2}) we have:
\begin{equation}
\label{clusterseteq3} \{C_{\NN}(v): v \in V(\NN)\} = \{C_{N}(v): v \in \VV(N)\},
\end{equation}
and thus the LHS of Eqn. (\ref{clusterseteq3})   is a hierarchy if and only if the RHS is. 
Finally, since $\NN$ is a normal network, it is a tree if and only if its cluster set
is a hierarchy, as required.
\hfill$\Box$

\blue{
\section{Algorithmic details}
The normalisation algorithm implemented in {\em PhyloSketch} proceeds as follows. 
First, for each vertex of the input network, we recursively compute the set of all lowest stable ancestors (LSA). To accomplish this, in a post-order traversal we label each vertex by the set of all descendant vertices. Then, for each vertex $v$, in a pre-order traversal we determine the first vertex for which two (or more) different children are labeled by $v$ as  a descendant.}

\blue{Next, in a post-order traversal, we determine the set of visible vertices by determining every vertex $v$ that is either an LSA, or is the only parent of a visible vertex or leaf. In addition, we compute the set of leaves as all vertices with out-degree 0.
Then, in a simple post-order traversal, we compute a mapping that maps each vertex $v$ to the set of all visible or leaf vertices below $v$.}

\blue{The normalisation network is then constructed as follows: For each original vertex $v$ that is either visible or a leaf, we create a new vertex $n(v)$. For each original vertex $v$ and for each visible or leaf vertex $w$ that lies below $v$, we create an edge from $n(v)$ to $n(w)$ in the new graph.
Finally, we perform transitive edge reduction on the new graph.}

\blue{In summary, the algorithm proceeds in a fixed number of traversals of the graph, each taking polynomial time in the number of vertices and edges. Transitive edge reduction takes at most cubic time (as a function of the number of vertices in the new graph) and so the algorithm runs in polynomial time.}

\blue{To obtain an estimate of running time, we generated random reticulated networks of increasing size and calculated the wall-clock time required on a laptop to compute the associated normalisation network.  For four such generated networks: $N_1$ (with 13,480 vertices, 17,564 edges and 2,329 reticulations), $N_2$ (27,316 vertices, 35,878 edges and 4,776 reticulations), $N_3$ (with 68,693 vertices, 90,459 edges and 12,020 reticulations) and $N_4$ (37,086 vertices, 180,466 edges and 23,982 reticulations), the time required to compute the normalisation network was 1, 3, 22 and 79 seconds, respectively.}

\end{document}